\documentclass[doublecol]{epl2} 

\usepackage{hyperref}
\usepackage{graphicx}

\usepackage{color}
\usepackage[normalem]{ulem} 

\usepackage{xcolor}
\definecolor{shadecolor}{RGB}{255,255,255}

\newcommand\omegad{\omega_{\rm d}}

\renewcommand\phi{\varphi}
\renewcommand\rho{\varrho}
\newcommand\omegac{\omega_{\rm c}}

\definecolor{DarkGreen}{rgb}{0,0.4,0}

\title{Multiple perfectly-transmitting states of a single-level at strong coupling}

\author{
\'Etienne Jussiau\footnote{\  ejussiau@UR.Rochester.edu}\inst{1,2} \and 
Robert S. Whitney\footnote{\ robert.whitney@grenoble.cnrs.fr}\inst{2}}
\shortauthor{\'E. Jussiau and R.S. Whitney}

\institute{                    
 \inst{1}Department of Physics and Astronomy, University of Rochester, Rochester, NY 14627, USA\\
 \inst{2} Laboratoire de Physique et Mod\'elisation des Milieux Condens\'es, Universit\'e Grenoble Alpes and CNRS,\\
 \phantom{\inst{2}} 25 Avenue des Martyrs, BP 166, 38042 Grenoble, France.
 }


\abstract{
We study transport through a single-level system placed between two reservoirs with band-structure,
taking strong level-reservoir coupling of the order of the energy-scales of these band-structures.   
An exact solution in the absence of interactions gives the nonlinear Lamb shift.
As expected, this moves the perfectly-transmitting state (the reservoir state that flows through the single-level without reflection), and can even turn it into a bound-state.
However, here we show that it can also create additional pairs of perfectly-transmitting states at other energies, when the coupling exceeds critical values. 
Then the single-level's transmission function resembles that of a multi-level system.
Even when the discrete level is outside the reservoirs' bands,
additional perfectly-transmitting states can appear inside the band when the coupling exceeds a critical value.  We propose observing the bosonic version of this in microwave cavities,
and the fermionic version in the conductance of a quantum dot coupled to 1D or 2D reservoirs.
}

\begin{document}

\renewcommand\thispagestyle{eplplain}

\maketitle


\vskip -5truemm  
\noindent\colorbox{shadecolor}
{\parbox{\dimexpr\columnwidth-2\fboxsep\relax}{$\phantom{|}$}}
\vskip -12truemm
$\ $

\section{Introduction}
There is currently great interest in strong coupling between small systems and reservoirs
\cite{Apertet2012Mar,Topp2015,Katz2016May,Strasberg2017Jun,Strasberg2018May,Whitney2018Aug,Dou2018Oct,Seifert-PRL2016,Jussiau2019Sep}, 
for both practical and fundamental  reasons.
It is of practical interest in electron nanostructures; for example for lower resistance electronics (resistance being inversely proportional to coupling strength), and more powerful nanoscale thermoelectric heat engines and refrigerators \cite{ReviewBCSW}. 
It is of fundamental interest, because it leads to physics 
which one could not guess from either the small system's properties or the reservoir's properties.
In such cases, the reservoir can induce long time correlations in the small system's evolution, making it non-Markovian,
and thus challenging to model. 

\begin{figure}
\centerline{\includegraphics[width=0.9\columnwidth]{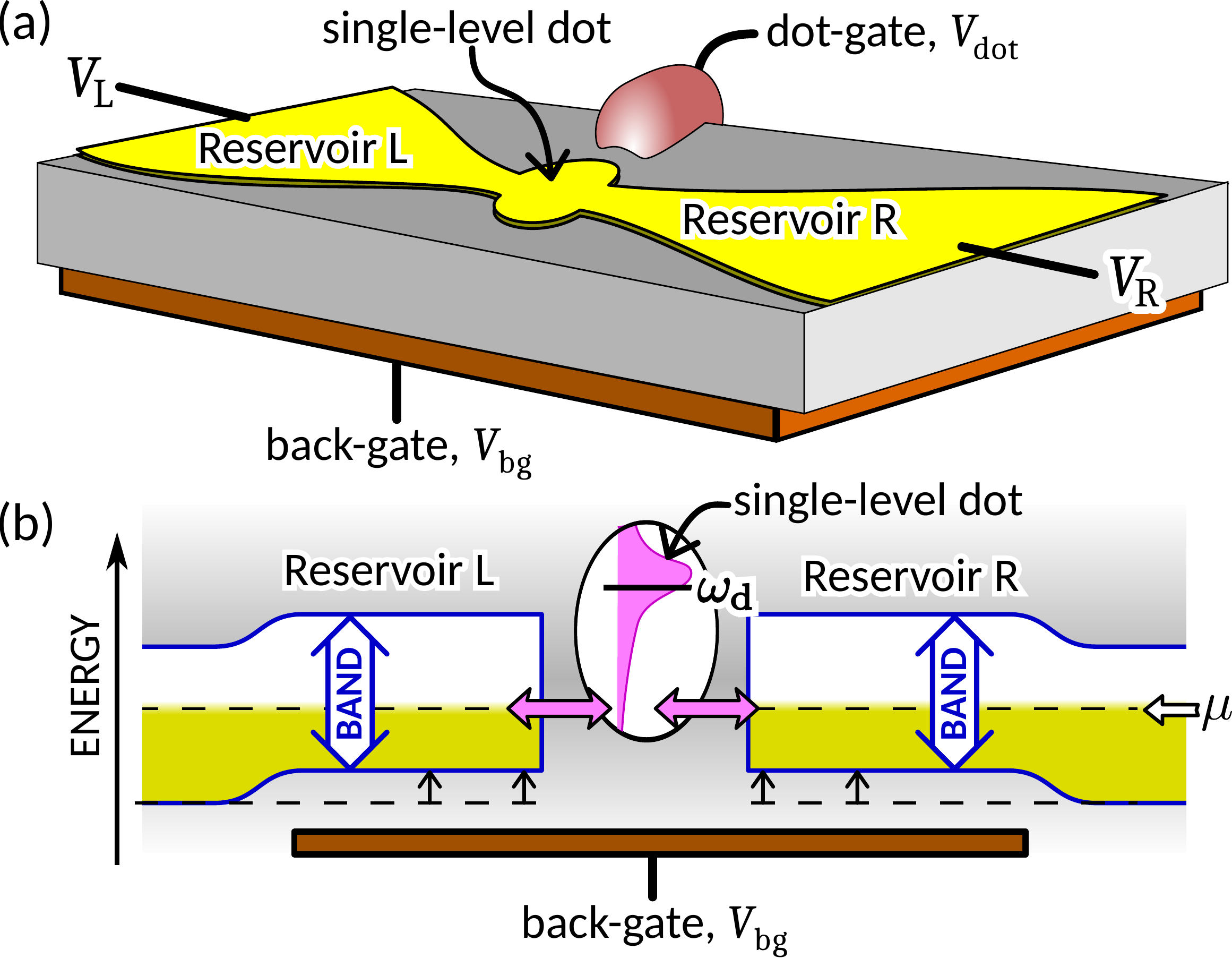}} 
\caption{(a) A single-level quantum dot between two 2-dimensional electronic reservoirs; the reservoirs could equally be 1D wires.
(b) The back-gate is used to shift the reservoir's band with respect to the electro-chemical potential, $\mu$. 
The dot-gate controls the energy of the dot-level, $\omegad$.
The purple Lorentzian is the golden-rule picture, which {\it cannot} explain the perfectly-transmitting states that we find.
}
\label{Fig-System}
\end{figure}

\begin{figure}
\centerline{\includegraphics[width=0.9\columnwidth]{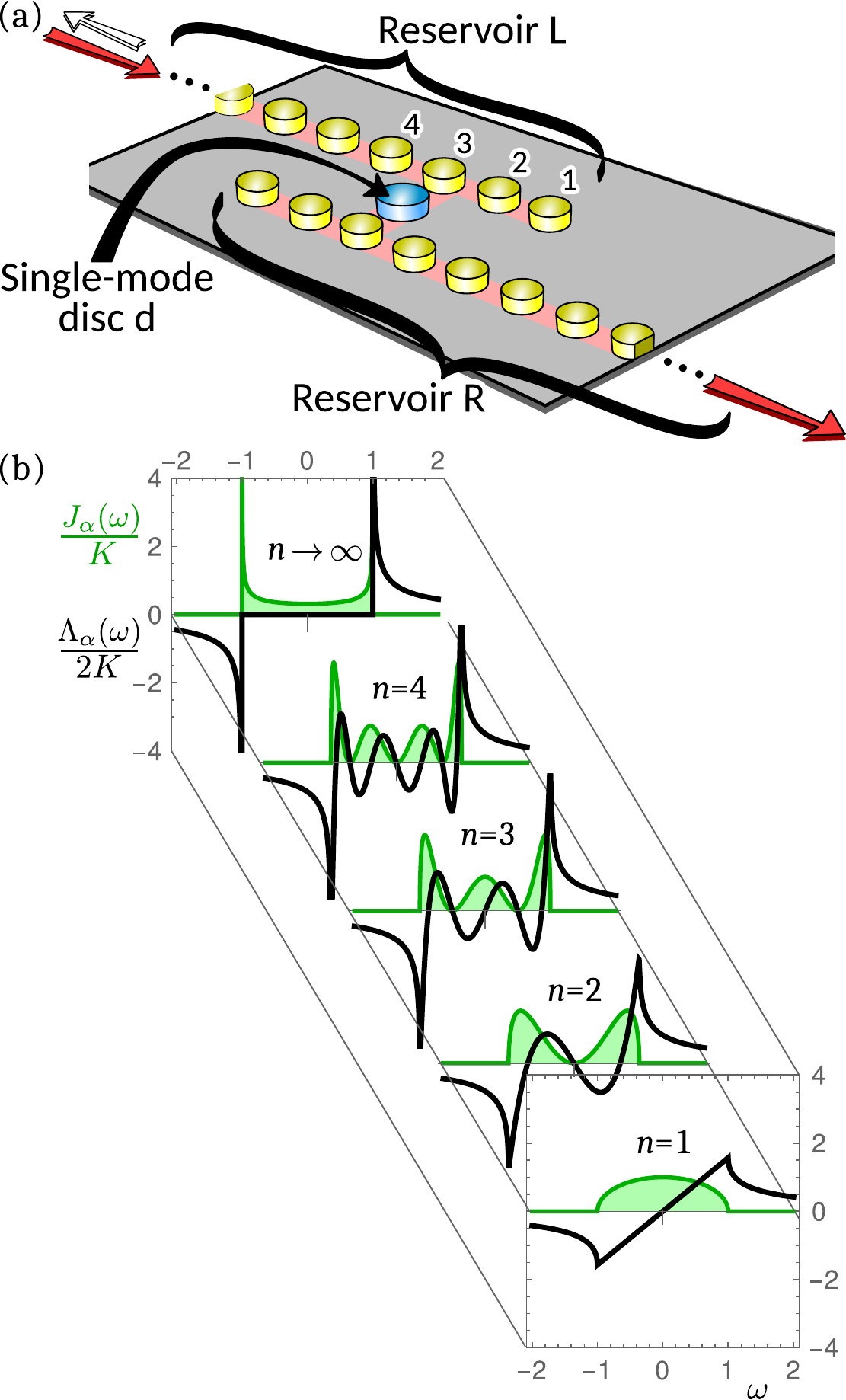}}
\caption{(a) A microwave system in which the single-level is the single-mode of disc d, while 
each reservoir is a chain of single-mode dielectric discs (yellow),
with weak coupling to the nearest neighbour in the chain (indicated by pink bands).  
Disc d's single-mode is at an energy which differs by $\omegad$ 
from those in the reservoir chains,
and it couples to the $n$th disc in each reservoir (the sketch shows $n=3$) with
coupling strength $K$. 
(b) 
The spectral function and Lamb shift when disc d is coupled to the $n$th disc in the chain, plotted as $J_\alpha(\omega)/K$ in green and $\Lambda_\alpha(\omega)/(2K)$ in black
(the factor of 2 is to make $J_\alpha(\omega)$ and $\Lambda_\alpha(\omega)$ appear clearly in the same plot).
}
\label{Fig-microwaves}
\end{figure}

In this work, we show that strong coupling has a striking effect on {\it perfectly-transmitting states}.
These are reservoir states that exhibit perfect (reflectionless) flow from one reservoir to another through 
a small system.
We consider the small system to be a single-level system (or single-mode cavity) as in Figs.~\ref{Fig-System} \& \ref{Fig-microwaves}.
We show that the perfectly-transmitting states are intimately related
to {\it bound-states},
despite having the opposite properties (bound states carry no steady-state flow between reservoirs\footnote{We consider the dc (zero-frequency) component of the current, established at long times after turning on the coupling.  This clarification is important because bound-states can induce finite-frequency oscillations which survive in the long-time limit, if the coupling is turned on non-adiabatically \cite{Stefanucci2007May,Jussiau2019Sep}.}).
Such bound-states are a known (but intriguing) consequence of strong-coupling to a reservoir with a band-gap,
giving non-Markovian dynamics to the single-level system \cite{Zhang2012}. 
Then a particle placed in the single-level system never fully decays into the continuum even when it has the energy to do so \cite{John1990, John1991,John1994,Kofman1994,Angelakis2004,Chang2018,Xiong2015Aug}.
Bound-states were first predicted for an impurity in a superconductor \cite{Shiba1973Jul}, or an electronic tight-binding model
\cite{BookMahan}.
They were studied for the Wigner-Weisskopf
model of an atomic level coupled to a photonic vacuum with band-structure \cite{John1990, John1991,John1994,Kofman1994},
reviewed in Refs.~\cite{Angelakis2004,Chang2018}, 
with generalizations to a finite-temperature Fano-Anderson model
\cite{Zhang2012,Lei2012May,Xiong2015Aug,Ali2015Dec,Ali2017Mar}.
Their properties were explored in models of quantum dots coupled 
to electronic reservoirs
\cite{Maciejko2006Aug,Dhar2006Feb,Stefanucci2007May,Jin2010Aug,Xiong2015Aug,Yang2015Oct,Tu2016Mar,Zhang2012,Engelhardt2016,Lin2016,Basko2017}.
Recent experiments  probed them in NV centres in a waveguide \cite{Liu2017},
and for matter waves in ultracold atoms \cite{Krinner2018Jul}.
While such bound-states have rich physics, we show that the physics of 
{\it perfectly-transmitting states} is similar, but richer.

Our main result is that additional perfectly-transmitting states
appear via transitions that occur when the coupling exceeds critical values.
Then a single-level's transmission function resembles that of a multi-level system.
The most striking example is when the level's energy is outside the reservoirs' band.
Once the coupling exceeds a critical value,  additional perfectly-transmitting states appear within the band, 
ensuring perfect (reflectionless) flow between the reservoirs, 
even though the flow is through a level at an energy outside the band.

\section{Perfectly-transmitting states in golden-rule}
At weak-coupling, the perfectly-transmitting state of a single-level system
is described by Fermi's golden rule \cite{Cohen-TannoudjiGoldenRule}.
For extremely weak coupling, resonant transmission happens at the energy of this level, so there is a perfectly-transmitting state at this energy. The golden rule then implies that as the coupling increases,
 this state is broadened and Lamb shifted.
 However, this argument does not predict more than one perfectly-transmitting state per discrete-level, unlike our exact calculation.

\begin{figure*}
\includegraphics[width=\linewidth]{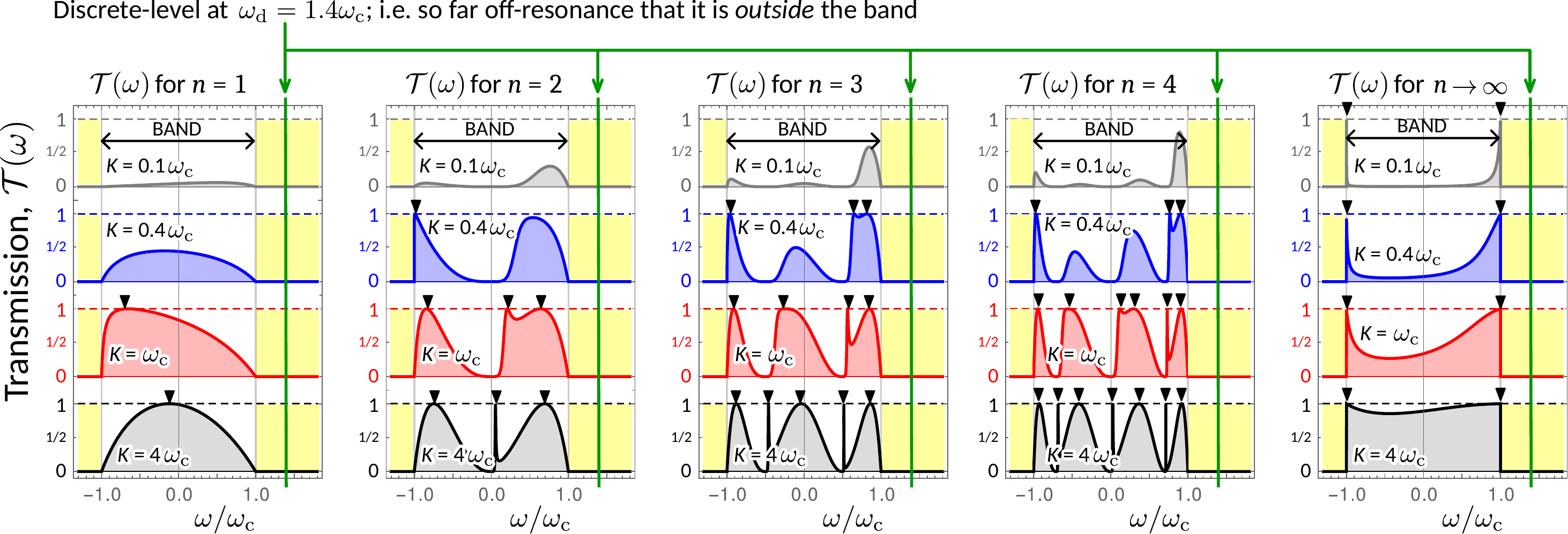}
\caption{
Transmission functions when the single-level is outside the reservoir's band at $\omega_{\rm d}=1.4\omegac$.
Despite this, there are perfectly-transmitting states inside the band (indicated by inverted black triangles),
at strong enough coupling. The plots are for systems with the spectral functions, 
$J_{\rm L}(\omega)=J_{\rm R}(\omega)$ given by Eq.~(\ref{Eq:J-1D-site-n}), as for the 
system in Fig.~\ref{Fig-microwaves}, with $n=1,2,3,4, \infty$. Note that the bound-states do not appear here because they do not contribute to the transmission; when they are present, they are outside the band (sitting in the yellow zones).
}
\label{Fig-transmission1D}
\end{figure*}

If the single-level has an energy outside the reservoir's band, then 
this golden-rule argument would predict 
that the level is Lamb shifted away from the band 
(level-repulsion between the level and the reservoir modes), while it broadens into a Lorentzian.
This would suggest that transmission between reservoirs only occurs through the Lorentzian's tail (see  Fig.~\ref{Fig-System}b). 
Hence, this golden-rule argument cannot explain the perfectly-transmitting states
that appear in our exact calculation, see Fig.~\ref{Fig-transmission1D}.

This golden-rule argument fails due to its neglect of the energy dependence of the Lamb shift,
induced by the reservoirs' band-structure.
It is the nonlinearity of this energy dependence that gives rise to transitions at which new perfectly-transmitting states appear.

\section{Hamiltonian and transmission function}
We consider the Fano-Anderson 
model\cite{fano1961,Anderson1961,BookMahan,BookCohen-Tannoudji} 
for a single-level system coupled to two reservoirs; left (L) and right (R).
This model applies to electrons (Fig.~\ref{Fig-System}) in the absence of spin and
interaction effects\footnote{The model applies to quantum dots with on-site Coulomb interactions, so long as  the magnetic field is strong enough that the spin-state with higher energy is always empty.}, or with interactions treated in the Hartree approximation \cite{Anderson1961}.
It also applies to photons\cite{fano1961,BookCohen-Tannoudji} (Fig.~\ref{Fig-microwaves}a), 
or other non-interacting bosons \cite{Engelhardt2016}. 
The model's Hamiltonian is 
\begin{equation}
\hat H=\omegad\hat d^\dagger\hat d+\sum_{\alpha,k}
\left(\, \omega_{\alpha k}\hat c_{\alpha k}^\dagger\hat c_{\alpha k} + g_{\alpha k}\hat d^\dagger\hat c_{\alpha k}+g_{\alpha k}^*\hat c_{\alpha k}^\dagger\hat d \,\right),
\label{H}
\end{equation}
where $\hat d^\dagger$ and $\hat d$ are creation and annihilation operators for the single-level system state,
while $\hat c_{\alpha k}^\dagger$ and $\hat c_{\alpha k}$ are creation and annihilation operators for
the $k$th mode of reservoir $\alpha\in\{{\rm L},{\rm R}\}$.
The discrete state has energy $\omega_{\rm d}$, and has a coupling $g_{\alpha k}$ to the $k$th mode in reservoir $\alpha$, which has energy $\omega_{\alpha k}$.   
The creation and annihilation operators are fermionic for electrons, while they are bosonic for photons.
In either case, each reservoir contains infinitely many modes described by continuous spectral functions,
\begin{equation}
J_\alpha(\omega)=\sum_k |g_{\alpha k}|^2\delta(\omega-\omega_{\alpha k}).
\label{J}
\end{equation}
Physically, this spectral function is equal to the reservoir's local density-of-states on its surface where 
it exchanges particles with the discrete-level, multiplied by the tunnelling amplitude, $|g_{\alpha k}|^2$. 
For convenience, we define a dot-reservoir coupling $K_\alpha$ as the typical magnitude of $J_\alpha(\omega)$ (its exact definition
will be given below for specific reservoir spectra).
We do \textit{not} take the wide-band limit, and instead consider the coupling $K_\alpha$ to be of the order of the band-width.
This typically requires $K_\alpha$ to be of similar magnitude to the inter-site coupling inside the reservoir.

An exact solution to the model defined by Eqs.~(\ref{H},\ref{J}) using non-equilibrium Green's functions \cite{Meir1992,Stefanucci2007May,BookRyndyk2009,Yang2015Oct} or simply using Heisenberg equations of motion \cite{Topp2015,Jussiau2019Sep},
gives the dc particle and energy currents (see footnote $^1$),
\begin{eqnarray}
j_{\rm dc}^{(\kappa)}&=&\int_{-\infty}^\infty\frac{{\rm d}\omega}{h}\ \omega^\kappa\ \mathcal T(\omega)(f_{\rm L}(\omega)-f_{\rm R}(\omega)),
\label{Eq:j_of_kappa}
\end{eqnarray}
where $f_\alpha(\omega)$ is reservoir $\alpha$'s distribution function, 
with $\kappa=0$ for particle current and $\kappa=1$ for energy current.
Hence, this exact calculation gives currents  that take the form of a Landauer formula\cite{Landauer1957,Landauer1970,Caroli1971,Meir1992}.
The transmission function, $\mathcal T(\omega)$, is the probability that a particle at energy $\omega$ transmits from one reservoir to the other through the single-level.
The exact solution gives \cite{BookRyndyk2009,Topp2015,Martensen2019Feb,Jussiau2019Sep},
\begin{equation}
\mathcal T(\omega)=\frac{4\pi^2 \, J_\mathrm L(\omega)\,J_\mathrm R(\omega)}{(\omega-\omegad-\Lambda(\omega))^2+\pi^2\big(J_\mathrm L(\omega)+J_\mathrm R(\omega)\big)^2}\, ,
\label{transmission}
\end{equation}
for all $\omega$ where $J_\mathrm{L,R}(\omega)\ne0$,
with $\mathcal T(\omega)=0$ otherwise. 
Eq.~(\ref{transmission}) is a distorted Lorentzian if $J_\alpha(\omega)$ and $\Lambda(\omega)$ only depend weakly on $\omega$, but we will show that it can take very different shapes for strong $\omega$ dependences. For any $\omega$-dependence, it obeys $0 \leq \mathcal T(\omega) \leq 1$ for all $\omega$.
Here $\Lambda(\omega)=\Lambda_\mathrm L(\omega)+\Lambda_\mathrm R(\omega)$ is the Lamb shift, 
given by the principal value integral
\begin{equation}
\Lambda_\alpha(\omega)= P\hskip-3.75mm\int {\rm d}\omega'\,\frac{J_\alpha(\omega')}{\omega-\omega'}\, .
\label{Lambda}
\end{equation}

For photonic reservoirs, ${\cal T}(\tilde\omega)$ is directly observable from Eq.~(\ref{Eq:j_of_kappa})
by injecting a monochromatic beam at frequency $\tilde\omega$  into the system; 
so $f_{\rm L}(\omega)\propto\delta(\omega-\tilde\omega)$ and $f_{\rm R}(\omega)=0$.
For electronic reservoirs, one measures this transmission function by going to low temperature
and small bias, where Eq.~(\ref{Eq:j_of_kappa}) 
gives the electrical current $I_{\rm dc}=ej_{\rm dc}^{(0)}= GV$
with conductance 
\begin{eqnarray}
G = e^2 \mathcal T (\mu) \big/ h \,,
\label{Eq:G}
\end{eqnarray}
where $\mu$ is the reservoirs' electrochemical potential.  
Then $\mathcal T (\omega)$ can be probed
by measuring $G$ while using a back-gate to move $\mu$ with respect to the reservoir bands \cite{Williams2007,Chen2010,Lemme2011} (see  Fig.~\ref{Fig-System}b). 
Note that to-date experiments that use the back-gate in this manner require that the reservoirs are one or two dimensional; nanowires, graphene, etc.

\section{Perfectly-transmitting states and bound states}
If the two reservoirs have the same spectral function, $J_{\rm L}(\omega)=J_{\rm R}(\omega)$,
 both the perfectly-transmitting states and the bound-states 
are given by the solutions of 
\begin{eqnarray}
\omega -\omegad = \Lambda(\omega).
\label{peak-condition}
\end{eqnarray}
The solutions with $\omega$s that fall in band-gaps (so $J_{\rm L}(\omega)=J_{\rm R}(\omega)=0$) are well known to be 
bound-states \cite{OLeary1990,BookMahan,BookCohen-Tannoudji}.
Here we see all other solutions are perfectly-transmitting states,
because they have ${\cal T}(\omega)=1$
from Eq.~(\ref{transmission}) with $J_{\rm L}(\omega)=J_{\rm R}(\omega)\neq 0$.
All such solutions can be found graphically by plotting $ \Lambda(\omega)$, and seeing where it intersects the line $(\omega-\omegad)$.

For any $J_\alpha(\omega)$, Eq.~(\ref{Lambda}) shows that $\Lambda(\omega)$ 
decreases monotonically with $\omega$ in any band gap, 
so  there is at most one bound state in any given band-gap \cite{Jussiau2019Sep}.
In contrast, $\Lambda(\omega)$ can be non-monotonic inside bands, 
see Figs.~\ref{Fig-microwaves}b and \ref{Fig-graphene-Bfield}a for examples. 
Then there can be multiple solutions of Eq.~(\ref{peak-condition}) inside a band, 
meaning multiple perfectly-transmitting states in that band.
At strong coupling, their number equals the number of zeros of $\Lambda(\omega)$ inside that band.  Often this is the maximum number of perfectly transmitting states, however
there are models with more such states at weak or intermediate coupling (see Appendix~C).

We concentrate on $J_{\rm L}(\omega)= J_{\rm R}(\omega)$ in this work.
However, we note that there are no perfectly-transmitting states for $J_{\rm L}(\omega)\neq J_{\rm R}(\omega)$.
Instead, solutions of Eq.~(\ref{peak-condition}) correspond to the maximum allowed transmission, 
$\mathcal{T}_{\rm max}(\omega) = 4J_{\rm L}(\omega)J_{\rm R}(\omega)\Big/\big(J_{\rm L}(\omega)+J_{\rm R}(\omega)\big)^2$; that is to say that
no choice of  $\omegad$ that will give a larger ${\cal T}(\omega)$ at this $\omega$.

\begin{figure}
\includegraphics[width=0.95\linewidth]{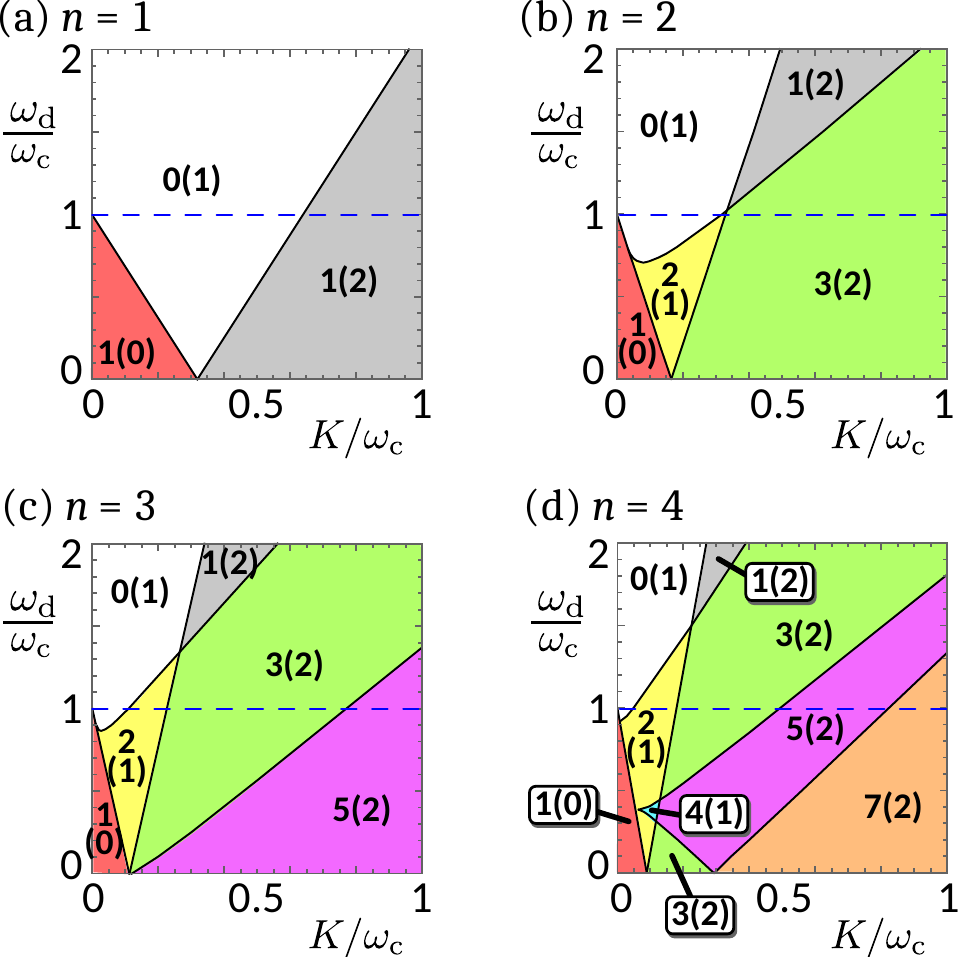}
\caption{
Phase diagrams
for the system in Fig.~\ref{Fig-microwaves}a when disc d is coupled to the $n$th disc in each 1D reservoir. Each ``phase'' labelled $k\,(l)$ where $k$ is the number of perfectly-transmitting states
and $l$ is the number of bound-states.
}
\label{Fig-PhaseDiagram}
\end{figure}

\section{Phase-diagrams}
By counting the solutions of the Eq.~(\ref{peak-condition}),
we can construct ``phase diagrams'' of the number of perfectly-transmitting states and bound states as a function of the discrete level's energy, $\omegad$, and the dot-reservoir coupling, $K_\alpha$, as in Figs.~\ref{Fig-PhaseDiagram} and \ref{Fig-graphene-Bfield}c.
Since $\Lambda(\omega)$ is a continuous function of $\omega$, new solutions of Eq.~(\ref{peak-condition}) appear in pairs.
At the same time all the solutions move with coupling.
Hence, there are two main types of transitions.
Transitions of type (i) are where a perfectly-transmitting state moves out of the continuum to become a bound state,
as discussed in Ref.~\cite{Jussiau2019Sep}.
Transitions of type (ii) are those where an additional pair of perfectly-transmitting states is added.
In this case, as the coupling approaches the phase boundary, a peak grows to form a perfectly-transmitting state at the phase boundary, which splits into two such states upon passing into the new phase.

There is also a non-generic transition, which we call type (iii), in which types (i) and (ii) occur together; i.e.~a pair of states appear at the band edge, but one immediately moves out of the band to become a bound-state while the other moves into the band as a perfectly-transmitting state.  
It is non-generic because it only happens when $\Lambda(\omega)$ has a cusp at the band-edge, 
and its first and second derivative have the same sign on the band side of the cusp.
However, this is the case for all the models in Fig.~\ref{Fig-microwaves}b, so type (iii) transitions occur there.

\section{Perfect transmission with level outside band}
If $\omegad$ is not within a band, then there are no perfectly-transmitting states
at weak-coupling (assuming $J_\alpha(\omega)$ and $\Lambda(\omega)$ do not diverge).
However, as the coupling is increased, transitions of type (ii) or (iii) can occur which generate
one or more perfectly-transmitting states inside the band.
Transitions from white to yellow in Fig.~\ref{Fig-PhaseDiagram} are type (ii), while those from white to grey are type (iii). 

In addition, for $J_{\rm L}(\omega)=J_{\rm R}(\omega)$ 
there are perfectly-transmitting states at any $\omega$ for which
$J_{\rm L}(\omega)$ diverges, if it diverges faster than $\Lambda(\omega)$. 
Then Eq.~(\ref{transmission}) gives ${\cal T}(\omega)=1$ at this $\omega$ for all $\omegad$ and all $K_\alpha$, even $K_\alpha\to0$ \cite{Jussiau2019Sep}; see e.g.\
${\cal T}(\omega)$ at the band-edge for $n\to \infty$ in Fig.~\ref{Fig-transmission1D}, due to  $J^{(n\to\infty)}_\alpha(\omega)$ having a square-root divergence there.

\section{Microwave tight-binding reservoirs}
We imagine microwave experiments on tight-binding 
Hamiltonians like in Refs.~\cite{Laurent2007Dec,Kuhl2010Sep,Barkhofen2013Jan}, with
high-refractive index dielectric discs arranged as in Fig.~\ref{Fig-microwaves}a, 
and sandwiched between two metallic plates.
Each centimetre-sized disc supports a single GigaHertz mode, whose energy is tuned
by changing the disc size.
The discs couple to each other via evanescent waves in the air between the metallic plates, 
so the tunnel coupling can be tuned by changing the distance between discs.  
The two one-dimensional chains of identical discs play the role of the reservoirs,
and they are both coupled to a different-sized disc (disc d), 
whose resonance is detuned from the others
by an energy $\omegad$,
as in Ref.~\cite{Longhi2007May}, but with two reservoirs instead of one.
We calculate the transmission from one chain to the other, ${\cal T}(\omega)$, to find the perfectly-transmitting states.

We define  $\omega=E-E_0$, where $E$ is the injected microwave's energy,
and $E_0$ is the energy of the middle of the reservoir's band (given by the energy of the mode of a reservoir disc in isolation).
Then if disc d couples to disc $n$ in reservoir $\alpha \in\{{\rm  L,R}\}$, 
the reservoir's spectral function is proportional to the local density-of-states at disc $n$ in the reservoir \cite{Hoekstra1988Nov}.
As shown in Fig.~\ref{Fig-microwaves}b, these are \cite{Longhi2007May}
\begin{eqnarray}
J^{(n)}_\alpha (\omega)
&=& \left\{ \begin{array}{cl}
{\displaystyle {K \sin^2(n \phi_y) \over \sqrt{1- y^2}} } & \hbox{for } |y| \leq 1\,, \\
0 &  \hbox{for } |y| \geq 1 \,, \\
\end{array}\right.
\label{Eq:J-1D-site-n}
\end{eqnarray}
where  $y=\omega/\omegac$ and $\phi_y=\arccos[y]$,
which means \cite{Longhi2007May}
\begin{eqnarray}
\Lambda^{(n)}_\alpha (\omega)
\!\!\! &=& \!\!\!
\left\{\begin{array}{cl}
{\displaystyle {\pi K \,\sin[2 n \phi_y ] \over 2 \sqrt{1-y^2} } }&
\hbox{ for }  |y|\leq 1 \,,
\\
{\displaystyle \pi K \ {1-(y{\cal A}(y)-y)^{2n} \over 2y{\cal A}(y)} }
&
\hbox{ for }  |y|\geq 1 \,,
\end{array}\right.
\label{Eq:Lambda-1D-site-n}
\end{eqnarray}
where ${\cal A}(y)=\sqrt{1-y^{-2}}$. 
For $n\to \infty$, one has
$J^{(n\to\infty)}_\alpha (\omega)={\rm Re}\big[K \big/2\sqrt{1-y^2}\,\big]$.
Hence, $\Lambda^{(n\to\infty)}_\alpha(\omega)= {\rm Re}\big[\pi K\big/\big(2y{\cal A}(y)\big)\big]$, which is zero inside the band.

The transmission function, $\mathcal T(\omega)$, is then given by Eq.~(\ref{transmission}).
For all $n<\infty$, if the energy of disk d is outside the band ($|\omegad| >\omegac$), then the transmission at all $\omega$
is very small at weak-coupling, 
see the grey curves in Fig.~\ref{Fig-transmission1D}. 
It is surprising to find that once the coupling exceeds a critical value, then ${\cal T}(\omega)=1$ at certain values of $\omega$ inside the band
(marked by inverted black triangles in  Fig.~\ref{Fig-transmission1D}). 
This means that microwaves injected into reservoir L at these energies will be perfectly transmitted into reservoir R
(without any reflection), despite passing through disk d whose energy is not even within the band.

Fig.~\ref{Fig-PhaseDiagram} shows the phase diagram of perfectly-transmitting states for various $n$.
At strong coupling, there are $(2n\!-\!1)$ such
states (in addition to two bound-states outside the band)  whether disc d's energy is in the band or not, because Eq.~(\ref{Eq:Lambda-1D-site-n}) has $(2n\!-\!1)$ zeros in the band.

As the coupling grows, every alternate perfectly transmitting state broadens to a width of the order of the distance between peaks,
however the other peaks move towards energies where $J_\alpha(\omega)$ vanishes,
and so become extremely sharp. 
If one fine-tunes the dot energy $\omegad$ one can align one of these states exactly at an energy where $J_\alpha(\omega)$ vanishes, at which point it becomes a ``bound state in the continuum''\cite{Hsu2016Jul} of the type discussed in Ref.~\cite{Longhi2007May}.
However, the perfectly-transmitting states are always present and do not require any such fine-tuning.

For $n\to\infty$, the peaks become so densely packed that the transmission becomes smooth,
and $J^{(n\to\infty)}_\alpha(\omega)$ has a square-root divergence at $\omega=\pm\omegac$.  
Then there are perfectly transmitting states at the band-edges\cite{Jussiau2019Sep} at all $K$.

\begin{figure}
\includegraphics[width=0.95\columnwidth]{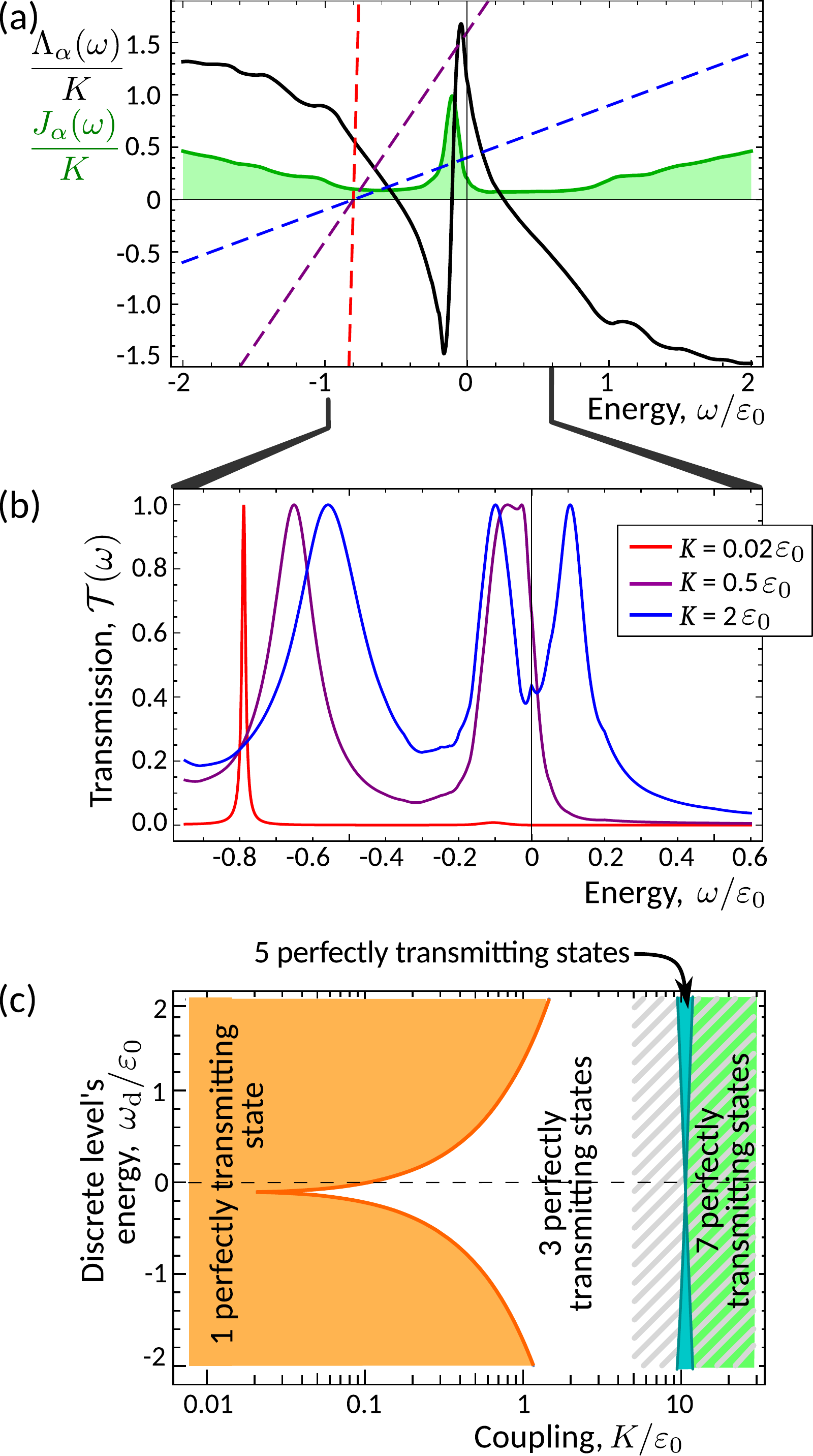} 
\caption{ 
(a) The spectral function $J_\alpha(\omega)/K$ (green curve) taken from Fig.~2c of Ref.~\cite{Li2013Apr}, and $\Lambda_\alpha(\omega)/K$ (black curve) that we calculate from it. The dashed lines are $(\omega-\omegad)/K$ for $\omegad/\varepsilon_0=-0.8$ and $K/\varepsilon_0=0.02,0.5,2$. Perfectly-transmitting states, given by Eq.~(\ref{peak-condition}), occur when the dashed lines intersect the black curve. 
(b) Plot of the transmission ${\cal T}(\omega)$ for $\omegad/\varepsilon_0=-0.8$. For small $K$, it is a single Lorentzian peak at $\omega=\omegad$, which broadens and moves as $K$ increases. A transition occurs (at $K/\varepsilon_0=0.45$) at which two more perfectly-transmitting states appear.
(c) Phase diagram showing the number of such states as a function of $\omegad$ and $K$.
}
\label{Fig-graphene-Bfield}
\end{figure}

\section{Graphene reservoirs}
Electronic systems rarely exhibit a single band well-separated from all others. Here we show we do not require this, 
the same physics can be seen when the reservoirs simply have a strong peak in their spectral function, $J_\alpha(\omega)$.
For this, we consider a single-level quantum dot between the zigzag edges of 
graphene reservoirs, under strong magnetic field. 
We take an experimental plot of $J_\alpha(\omega)$ from Ref.~\cite{Li2013Apr},  reproduced as
the green curve in Fig.~\ref{Fig-graphene-Bfield}a, which exhibits a strong peak. 
This curve comes from a STM measurement of the tunneling density-of-states of graphene's zigzag edge 
in the quantum Hall regime ($4\,$Tesla at $4.4\,$ Kelvin) \cite{Li2013Apr}.  
A similar peak was seen at the zigzag edge of a hexagonal lattice of microwave cavities \cite{Kuhl2010Sep}.
This peak is specific to the edge; the bulk's density of states is very different, and vanishes at the Dirac point ($\omega=0$ in  Fig.~\ref{Fig-graphene-Bfield}).
At sub-Kelvin temperatures, the dot's spin-splitting at $4\,$Tesla will be much bigger than temperature,
so we can take a single-level dot whose upper spin-state is empty at all times, making the dot well modelled
by Eq.~(\ref{H}).
The STM measurement gives us $J_\alpha(\omega)$ for $|\omega| <2\varepsilon_0$ 
where $\varepsilon_0= \sqrt{2e\hbar v_{\rm F}^2B}$ \cite{Li2013Apr}. 
We extrapolate 
this phenomenologically to $|\omega| >2\varepsilon_0$ as $\alpha |\omega|\exp\big[-|\omega|/\omegac\big]$
with $\omegac=10 \varepsilon_0$ and $\alpha$ matching the gradient at $\omega=\pm2\varepsilon_0$.
Numerically integrating over this $J_\alpha(\omega)$, gives the $\Lambda_\alpha(\omega)$ in Fig.~\ref{Fig-graphene-Bfield}a, 
and the dot's transmission in Fig.~\ref{Fig-graphene-Bfield}b,
to be observed experimentally with conductance measurements, see Eq. (\ref{Eq:G}).

For small $K$, there is one perfectly-transmitting state on-resonance (at the energy of the dot, $\omegad$). However as $K$ is increased, there is a transition to three such states.
The phase diagram in Fig.~\ref{Fig-graphene-Bfield}c shows up to seven such states for large coupling $K$.
However, the physics at large $K$ 
depends on $J_\alpha(\omega)$ at large $|\omega|$, and so depends on our chosen extrapolation beyond the experimental data at  $|\omega| >2\varepsilon_0$.
As such, the cross-hatched region of Fig.~\ref{Fig-graphene-Bfield}c may not agree with the experiments.

\section{Comparison with local density-of-states}
When $J_{\rm R}(\omega)=J_{\rm L}(\omega)$, the local density-of-states of the discrete-level is
${\cal T}(\omega) \big/\big(2\pi^2 J_{\rm L}(\omega)\big)$.
Hence we see that the peaks of the local density-of-states will not coincide with the perfectly-transmitting states,
unless $1/J_{\rm L}(\omega)$ has a very weak $\omega$ dependence at the $\omega$ in question. 
In particular, when $J_\alpha(\omega)$ has a square-root divergence (such as for $n\to \infty$ in Eq.~(\ref{Eq:J-1D-site-n}) or at the edge of the superconducting gap), the local density-of-states of the discrete-level vanishes, even though there is a perfectly-transmitting state there. 
Conversely,  the factor of $1\big/J_{\rm L}(\omega)$ in the local density-of-states means that it could have a peak at a value of $\omega$ which does not satisfy Eq.~(\ref{peak-condition}), if $J_{\rm L}(\omega)$'s gradient is large at this $\omega$.
Ref.~\cite{BookCohen-Tannoudji} used Eq.~(\ref{peak-condition}) as a heuristic method of finding  peaks in the local density-of-states.  The examples here show that this method is imprecise for such peaks (particularly at large $K$), 
when it is exact for perfectly-transmitting states at all $K$.

\section{Hand-waving picture}
To see why there are multiple perfectly-transmitting states, imagine replacing 
a peak in $J_\alpha(\omega)$ of height $K$ and width $\Omega$ with a Dirac $\delta$-function on a flat background, 
so $J_\alpha(\omega)\to K\Omega\delta(\omega-\omega_1)+K_0$.
Then $\Lambda(\omega)=2K\Omega\big/(\omega-\omega_1)$, 
giving three perfectly transmitting states at energies
$\omega_1$ and $\omega_\pm =\omega_1+{1 \over 2}\big[\omegad \pm \sqrt{\omegad^2 + 8K\Omega}\,\big]$. 
This can be explained by treating the $\delta$-peaks
as a single level in each reservoir. These fictitious levels at energy $\omega_1$ each have tunnel coupling  $\sqrt{K\Omega}$ to the discrete-level at $\omegad$. Solving this three-level system will give three eigenstates with energies; $\omega_1$ and $\omega_\pm$.
If $K_0$ is small enough that those three eigenstates have a golden-rule coupling to the remaining reservoir modes, it explains three perfectly-transmitting states at these energies.

Replacing finite-width peaks in $J_\alpha(\omega)$ with $\delta$-peaks
gets some features right for coupling much greater than the peak width, $K \gg \Omega$.
It correctly gives the solutions of Eq.~(\ref{peak-condition}) at $\omega \gg \omegac$
for $K/\omegac \to \infty$. It also correctly predicts that one solution will be 
at the centre of the spectral function's peak (i.e. at $\omega=\omega_1$) for  $K \gg \omegac,\omegad$.

This approximation can be improved by taking Lorentzians in place of $\delta$ functions.
Indeed, this is  a reasonable approximation for the peak in the experimental $J_\alpha(\omega)$ 
in Fig.~\ref{Fig-graphene-Bfield}.
Then
$J_\alpha(\omega)=K_\alpha\Omega^2 \big/\big( (\omega-\omega_1)^2+\Omega^2\big)$,
so Eq.~(\ref{Lambda}) gives the  Lamb shift  \cite{Topp2015,Martensen2019Feb}
\begin{eqnarray}
\Lambda_\alpha(\omega)
={\pi\,K_\alpha \Omega\ (\omega -\omega_1)\over (\omega-\omega_1)^2 +\Omega^2} \, .
\label{Lambda_lorentz}
\end{eqnarray}
Its qualitative features are a negative dip at $(\omega-\omega_1)\sim-\Omega$, 
growth through zero at $\omega\sim\omega_1$ and a positive peak 
at $(\omega-\omega_1)\sim\Omega$.
These qualitative features are the same for other-shaped peaks in $J_\alpha(\omega)$, such as in Fig.~\ref{Fig-microwaves}b.

A more sophisticated approach in the same spirit is the  ``reaction coordinate mapping'' \cite{Iles-Smith2014Sep,Nazir2018}, recently used\cite{Martensen2019Feb} on models like our Eq.~(\ref{H}).
Unfortunately, this mapping did not simplify the calculations presented in this work.

\section{Conclusions}
We show that a discrete-level strongly coupled to two reservoirs can exhibit multiple perfectly-transmitting states,
as if it were a multi-level system.  Even when the discrete-level is at an energy
outside the reservoirs' bands, such states can appear when the coupling exceed a critical value,
allowing perfect (reflectionless) flow between the two reservoirs.
We propose observing this in electronic and microwave systems.

This model is a good testing ground for quantum thermodynamics at strong coupling \cite{Ludovico2014Apr,Esposito2015Feb,Bruch2016Mar,Ludovico2016Jul,Whitney2018Aug,Seifert-PRL2016},
since the non-Markovian dynamics has such clear physical consequences for both bound states and perfectly-transmitting states.
Some of the consequences for thermoelectric effects have been studied \cite{Jussiau2019Sep}, 
but they merit further consideration.

\section{Acknowledgements}
We thank A.N.~Jordan, H.~Kurkjian, A.~Nazir and G.~Shaller for useful discussions.
We acknowledge the support of the French research program ANR-15-IDEX-02, via the Universit\'e Grenoble Alpes' QuEnG project.


\bibliographystyle{eplbib}   
\bibliography{ref}


\newpage

\newpage

\centerline{{\bf SUPPLEMENTARY MATERIAL}}
\vskip 5mm

\section{Appendix A: Simpler form of Eqs.~(\ref{Eq:J-1D-site-n},\ref{Eq:Lambda-1D-site-n})}
The spectral function, $J_\alpha^{(n)}(\omega)$, and the associated Lamb shift $\Lambda_\alpha^{(n)}(\omega)$
for the $n$th site in a one dimensional chain are given in compact forms by 
Eq.~(\ref{Eq:J-1D-site-n},\ref{Eq:Lambda-1D-site-n}), where we recall that $y\equiv\omega/\omegac$.
However, these are not the most transparent forms for energies inside the band, $|y| \leq 1$. 
There one can use the fact $\cos \phi_y = y$ with the multiple angle formulas in trigonometry to write the expressions in
terms of polynomials in $y$.
Then the spectral functions inside the band ($|y|\leq 1$) for small integer $n$ are 
\begin{eqnarray}
J^{(n=1)}_\alpha (\omega) 
&=& 
K_\alpha \sqrt{1-y^2}\, ,
\\
J^{(n=2)}_\alpha (\omega) 
&=&
K_\alpha \sqrt{1-y^2} \ \ 4y^2\, ,
\\
J^{(n=3)}_\alpha (\omega) 
&=&
K_\alpha \sqrt{1-y^2} \ \ \big(1-4y^2\big)^2\, ,
\\
J^{(n=4)}_\alpha (\omega) 
&=&
K_\alpha \sqrt{1-y^2} \ \ 16y^2 \big(1-2y^2\big)^2\, ,
\end{eqnarray}
with $y\equiv\omega/\omegac$.
We see that the second factor is an $(n-1)$-degree polynomial of $y^2$; so it becomes increasingly unpleasant to handle as $n$ grows.
Similarly, the Lamb shifts inside the band ($|y|\leq 1$) for these integer $n$ are 
\begin{eqnarray}
\Lambda_\alpha^{(n=1)}(\omega)
&=& \pi K_\alpha \,y \, ,
\\
\Lambda_\alpha^{(n=2)}(\omega)
&=&  -2\pi K_\alpha \,y\big(1- 2y^2 \big) \, ,
\\
\Lambda_\alpha^{(n=3)}(\omega)
&=&  \pi K_\alpha \,y\big( 3-16 y^2 + 16 y^4\big) \, ,
\\
\Lambda_\alpha^{(n=4)}(\omega)
&=&  -4\pi K_\alpha \,y\big(1-2 y^2 \big)\big( 1-8 y^2 + 8 y^4\big) \, .\qquad
\end{eqnarray}
We see that this is $y$ times an $(n-1)$-degree polynomial of $y^2$, so it is always odd in $y$.
These polynomials have $(2n-1)$ zeros in the window $|y|\leq 1$, as is most easily seen by counting the zeros of $\sin[2n\phi_y]$ in Eq.~(\ref{Eq:Lambda-1D-site-n}).

\section{Appendix B: Other simple examples of $\Lambda(\omega)$}
\label{append-s-examples}

\begin{figure}
\centerline{\includegraphics[width=0.8\linewidth]{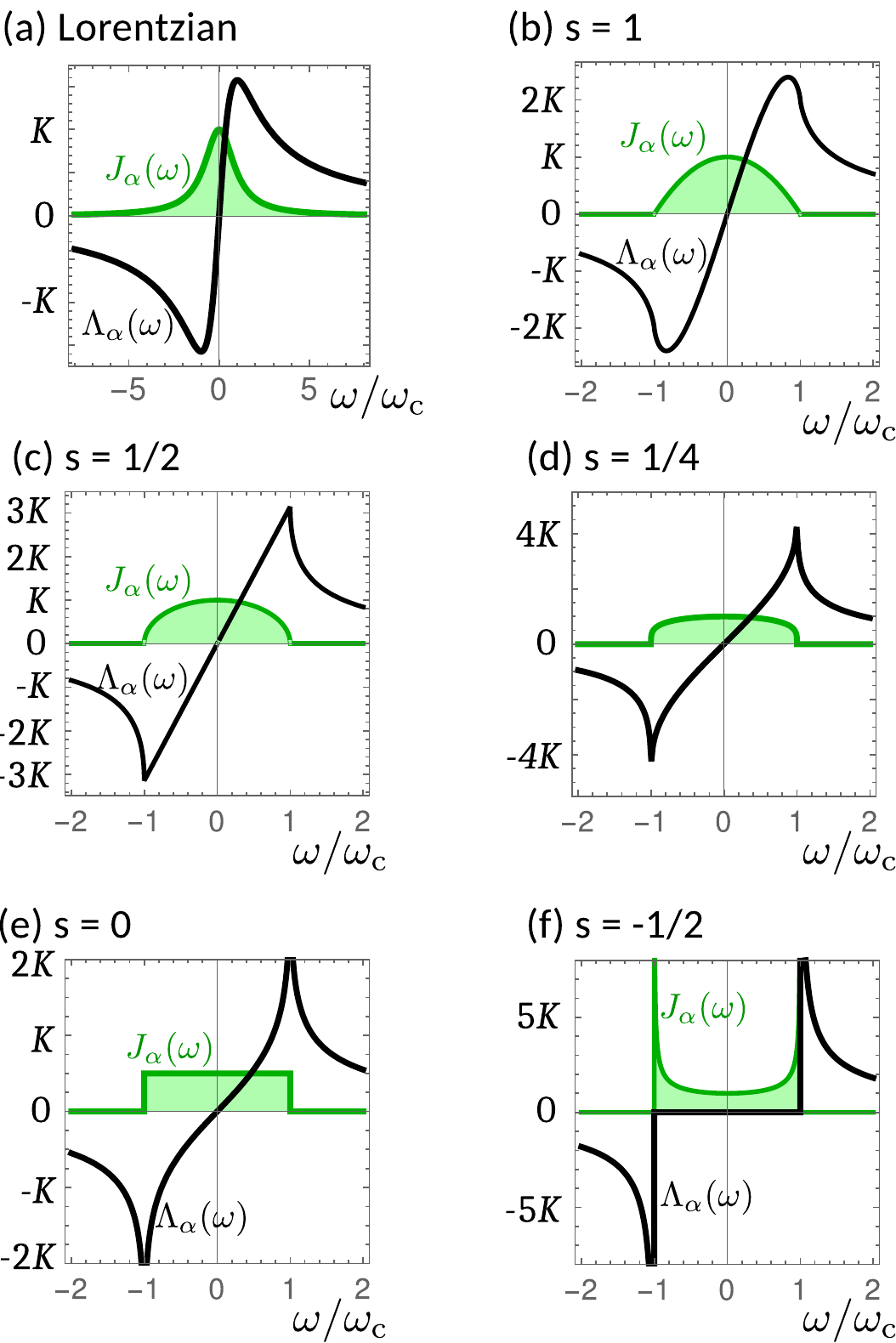}}
\caption{
Plots of various spectral functions, $J_\alpha(\omega)$ (green curve) and the associated Lamb shift, $\Lambda(\omega)$ (black curve). 
In (a) we plot a Lorentzian  $J_\alpha(\omega)$, 
with the $\Lambda_\alpha(\omega)$ in Eq.~(\ref{Lambda_lorentz}). 
In (b) to (f) we plot $J_\alpha(\omega)$ 
in Eq.~(\ref{J_s}) for different values of the spectral exponent, $s$, with the  
$\Lambda_\alpha(\omega)$ for most given in Eq.~(\ref{Lambda_s}). 
}
\label{Fig-Js_and_Lambdas}
\end{figure}

Phenomenologically, the simplest model of a single band is to assume its spectral function
takes the form
\begin{equation}
J_\alpha(\omega)=
\left\{\begin{array}{cl}
\ K \left(1-(\omega/\omegac)^2 \right)^s &
\mbox{for $|\omega| \leq \omegac,$}\\
\ 0&\mbox{for $|\omega| > \omegac,$}
\end{array}\right.
\label{J_s}
\end{equation}
where $s$ determines how the spectral function behaves at the band edges.
For certain values of $s$, one can directly evaluate
the integral in Eq.~(\ref{Lambda}).
Examples are
\begin{eqnarray}
\Lambda_\alpha(\omega)
\!\!\!\!&=& \!\!\!\!\left\{\!
\begin{array}{ll}
{\displaystyle K \left( {\omegac^2-\omega^2 \over \omegac^2}\, \ln \left|{\omegac+\omega \over \omegac-\omega}\right| +{2\omega \over \omegac} \right) }
&\!\! \hbox{for } s=1 
\\
{\displaystyle {\pi K \omega \over \omegac} \left( 1 -
{\rm Re}\left[\sqrt{1-(\omegac/\omega)^2}\,\right]   \right)
\!\phantom{{\Big| \over \Big|}} } 
& \!\!\hbox{for } s={1\over 2} 
\\
\ {\displaystyle K\, \ln \left|{\omegac+\omega \over \omegac-\omega}\right|  \phantom{{\Big| \over \Big|}} }& \!\!\hbox{for } s=0 
\\
\ {\displaystyle {\pi K \omegac\over \omega}  \, {\rm Re}\left[ \big(1-(\omegac/\omega)^2\big)^{-1/2}\right] } 
& \hskip -4mm\hbox{for } s=-{1\over 2}
\end{array}\right.
\nonumber \\
\label{Lambda_s}
\end{eqnarray}
where the positive square-root is taken, and 
the real parts are zero for $|\omega|<\omegac$.
These are shown in Fig.~\ref{Fig-Js_and_Lambdas}.
The results for  $s=\pm1/2$ were given elsewhere \cite{BookMahan,Xiong2015Aug,Yang2015Oct},
however it is intriguing that the Lamb shift inside the band is strictly zero for $s=-1/2$.

\section{Appendix C: When the strongest coupling does not have the most perfectly-transmitting states}
The spectral functions treated in the body of this work have the maximum number of perfectly-transmitting states
when the discrete-level's coupling to the reservoir is strongest.  This is not always the case.
Here we mention situations with more perfectly-transmitting states at weak or intermediate coupling than at strong coupling.

For spectral function, $J_\alpha(\omega)$, given by Eq.~(\ref{J_s}) with  $-1/2<s \leq 0$, one can see that $\Lambda(\omega)$ diverges at the band-edge,
and its gradient inside the band is not a concave function of $\omega$. 
Hence within the band, there can be multiple intersections between $\Lambda(\omega)$ and the straight line $(\omega-\omegad)$, even at weak-coupling (small or vanishing $K$);
implying multiple solutions of Eq.~(\ref{peak-condition}) inside the band (i.e. perfectly-transmitting states) when the coupling goes to zero, $K\to 0$.
This can be seen for $s=0$ in Fig.~\ref{Fig-Js_and_Lambdas}e, where a straight line will always intersect  $\Lambda(\omega)$
twice outside the band (i.e. two bound states), 
and will intersect $\Lambda(\omega)$ up to three times inside the band (i.e. up to three perfectly-transmitting states).
There are always three perfectly-transmitting states at weak-coupling, but only one perfectly-transmitting state above a certain critical coupling. 
The transition from three to one perfectly-transmitting states occurs when two such states converge and annihilate (a transition of type ii).

The fact that there are five solutions of Eq.~(\ref{peak-condition}) (two bound-states and three perfectly-transmitting states)
at arbitrarily small coupling
is a drastic indication that the golden-rule (which never predicts more than one such solution) fails even at vanishing coupling. A similar failure was already noted when there is a square-root divergence at the band-edge \cite{Jussiau2019Sep}, because  $J(\omega)$ exhibits a divergence.  
Now we can see that such a failure of golden-rule can also occur when 
$\Lambda(\omega)$ exhibit a divergence (even if $J(\omega)$ is convergent).

If there is no divergence in $J(\omega)$ or $\Lambda(\omega)$, then the golden rule will
work at small  coupling, and there will only be a single solution of Eq.~(\ref{peak-condition}) in the weak-coupling limit, $K\to 0$. This solution's energy will be extremely close to $\omegad$, so it will correspond to a bound-state if $\omegad$ is outside the band, and a perfectly-transmitting state if $\omegad$ is inside the band.
However, in this case, it is not difficult to find examples in which the maximum number of perfectly-transmitting states occurs at intermediate coupling.
Imagine taking almost any case where $\Lambda(\omega)$ does not diverge and is not a straight line inside the band,
such as $s=1$ or $s=1/4$ in Fig.~\ref{Fig-Js_and_Lambdas},
or indeed taking any $-1/2< s \leq 0$ with added rounding at the band-edge so divergences in $J(\omega)$ or $\Lambda(\omega)$ are replaced by large peaks.  Then it is easy to find situations with $\omegad$ close to the band centre in which there is;
\begin{itemize}
\item one perfectly-transmitting state with no bound states at small coupling,  
\item three perfectly-transmitting states and two bound states at intermediate coupling,
\item  one perfectly-transmitting state and  two bound states at large coupling.
\end{itemize}
So there are more such states at intermediate level-reservoir coupling than at large or small coupling.

\end{document}